\documentclass[aps,prl,twocolumn,showpacs,amsmath,amssymb]{revtex4}

\usepackage{epsfig}

\begin{document}
\title{Extracting the ground-state spin of a quantum dot from the conductance peaks in a parallel magnetic field at finite temperature} 
\author{Daniel Huertas-Hernando and Y. Alhassid}
\affiliation{Center for Theoretical Physics, Sloane Physics Laboratory, 
Yale University,New Haven, Connecticut 06520, USA.}

\date{\today}

\begin{abstract}
We derive a closed expression for the finite-temperature conductance of a Coulomb-blockade quantum dot in the presence of an exchange interaction and a parallel magnetic field. Parallel-field dependence
of Coulomb-blockade peak position has been used to determine experimentally the ground-state spin of quantum dots. We find
that for a realistic value of the exchange interaction, the peak motion can be significantly affected at temperatures as low as $kT \sim 0.1\,\Delta$, with $\Delta$ being the mean level spacing in the dot. 
This temperature effect can lead to misidentification of the ground-state spin when a level crossing occurs at low fields. We propose an improved method to determine unambiguously the ground-state spin. This method takes into account level crossings and temperature effects at a finite exchange interaction.
\end{abstract}

\maketitle

 The ground-state (GS) spin of a mesoscopic structure such as a quantum dot has been a topic of major interest \cite{Refspin}. In the absence of interactions, the total kinetic energy plus confining potential energy of the electrons is minimized in eigenstates with minimal spin. However, the exchange interaction is minimized in eigenstates of maximal spin.  The GS spin is therefore determined by the balance between these two competing energy terms.

The GS spin of a mesoscopic structure is difficult to measure.   In devices in which charging energy is important, Coulomb blockade (CB) serves as a useful tool to explore GS spin \cite{RefRMPReview}. In particular, the dependence of a CB   
peak position on a parallel magnetic field has been used to determine the GS spin in quantum dots \cite{RefCMMarcus,RefSiQD's1,RefQD,RefJFolk}, metallic nanoparticles 
\cite{SMG} and carbon nanotubes \cite{SCT}.  At temperatures $k T$ that are much smaller than the single-particle mean level spacing  $\Delta$ of the dot, the conductance peak position is determined by the change in the GS energy as an electron is added to the dot.  The Zeeman coupling to an in-plane magnetic field $B_\parallel$ is described by $g \mu_B B_\parallel\hat S_\parallel$, where $g$ is the gyromagnetic factor (taken to be positive), $\mu_B$ is the Bohr magneton, and $\hat S_\parallel$ is the component of the total spin along the field. Thus, in the presence of a parallel field, the GS energy of the dot with GS spin $S$ is shifted down by an amount $g \mu_B B_\parallel S$, corresponding to the state of lowest magnetic quantum number $M=-S$. The CB peak position acquires a shift $- g\mu_B B_\parallel \Delta S$, where $\Delta S$ is the change in the GS spin  as the number of electrons increases from $N$ to $N+1$.  The CB peak positions are then expected to exhibit a certain pattern of $\mp g \mu_B/2$ slopes versus $B_\parallel$, describing the increase or decrease of the GS spin by $1/2$ \cite{RefJS}.  

    An excited state in a dot with energy $E^*$ (at zero magnetic field) and spin $S^* >S$ will acquire a larger Zeeman energy shift and will cross the GS level $E_{\rm GS}$ at a value of the parallel field given by $g\mu_B B_\parallel (S^*-S)= E^* - E_{\rm GS}$. If the magnetic field is further increased, the excited energy level will become the new GS for 
the dot, resulting in an abrupt change in the CB peak position slope. The new slope is still determined by the change of the dot's total spin but it now involves 
the spin of the corresponding zero-field excited state instead of the GS spin.  In the presence of an exchange interaction, higher spin states shift down in energy, and crossings are expected to occur at smaller values of $B_\parallel$. 

Slopes of $\mp g\mu_B/2$ and traces of crossings are clearly seen in the small Si dots of Ref.~\cite{RefSiQD's1} for small values of $B_\parallel$ ($g\mu_B B_\parallel/\Delta \alt 0.3$). However, for the weakly coupled dots of Ref.~\cite{RefCMMarcus}, the observed slopes of the peak spacings at small $B_\parallel$ are essentially flat (see Fig.~9), and were difficult to interpret. Similar flat slopes were seen in Ref. \cite{RefQD}. 

Here we investigate CB peak motion in quantum dots under an applied parallel magnetic field
and at finite temperature. Remarkably, we find that signatures of level crossings can be almost completely washed out at 
temperatures as low as $k T \sim 0.1\,\Delta $. In general, we observe that the peak positions become flat at sufficiently low $B_\parallel$ (for which the Zeeman energy $g \mu_B B_\parallel$ is below $kT$)~\cite{RefQD}. However, when a crossing occurs at small values of $B_\parallel$, the peak position remains flat up to the crossing point. These findings suggest a possible explanation of the results of Refs.~\cite{RefCMMarcus} and \cite{RefQD}, for which $kT \sim 0.15\,\Delta$ and $kT \sim 0.2\,\Delta$, respectively.  If the slopes beyond the flat section are used, the
GS total spin can be misidentified. We propose an improved method in which the CB
peak position and peak height data are both fitted to a two-transition model that includes an excited state in the dot with either $N$ or $N+1$ electrons. We show that this method can detect all relevant crossings at temperatures $kT \sim 0.1 - 0.15 \,\Delta$,
allowing for a correct assignment of GS spins in a systematic way. 

  We assume a quantum dot that is weakly coupled to leads, such that  $\Gamma \ll kT$ and $\Gamma \ll\Delta$, where $\Gamma$ is a typical tunneling width of an electron from the leads into the dot. In this limit we can use a rate-equations approach to calculate the linear conductance $G$ in the presence of residual interactions~\cite{RefTGY}. The results of Ref.~\cite{RefTGY} can be generalized to include a parallel magnetic field. 

A closed solution for the conductance is not always possible. An important case for which an explicit solution exists is the universal Hamiltonian, obtained for a chaotic or diffusive dot in the limit of a large Thouless conductance \cite{RefKurland,RefRewAL}. In the presence of an 
applied magnetic field $B_{\parallel}$ parallel to the plane of the dot, the Hamiltonian is $\hat{H} =\sum_\lambda \epsilon_\lambda \hat n_\lambda + e^2\hat{n}^{2}/2 C -J_S \mathbf{\hat S}^{2}+g\mu_B B_\parallel \mathbf{\hat S}_{\parallel}$, where
 $\epsilon_\lambda$ are spin-degenerate single-particle levels,  $C$ is  the dot's capacitance, $J_{S}$ is the exchange interaction constant, and ${\bf \hat S}$ is the total spin of the dot. The term $g\mu_B B_\parallel  S_\parallel$ describes the Zeeman coupling to the field $B_\parallel$, where $\hat S_\parallel$ is the total spin projection along the field direction. The orbital-level occupations $n_\lambda$, spin $S$ and spin projection $M$ along the field direction are all good quantum numbers, and the
eigenenergies are $\varepsilon _{\alpha S,M}^{(N)}=\sum_{\lambda}\epsilon _{\lambda } n_{\lambda }+e^{2}N^{2}/2C-J_{S}S(S+1)+g\mu_B B_\parallel M$. 

 Here we generalize the expression for the conductance $G$ in the vicinity of an $N \rightarrow N+1$ CB peak,  obtained in Ref.~\cite{RefThomas}, to the case of an applied parallel field. We find
\begin{equation}\label{cond}
 G = {e^2 \bar\Gamma \over 2\hbar kT}
\sum_{\lambda }\left( w_{\lambda }^{(0)}+w_{\lambda
}^{(1)}\right)g_{\lambda }\;,
\end{equation}
 where $\bar \Gamma$ is the average width of a level, $\textstyle{g_\lambda\! = 2 {\bar\Gamma}^{-1} \Gamma_\lambda^{\rm l} \Gamma_\lambda^{\rm r}/(\Gamma_\lambda^{\rm l} + \Gamma_\lambda^{\rm r}})$ are the single-level (dimensionless) conductances, and $\Gamma^{\rm l(r)}_\lambda$
 are the single-level tunneling widths to decay from level $\lambda$ to  the left (right) lead. 
The thermal weights $w_{\lambda }^{(0)}$ and $w_{\lambda }^{(1)}$
collect the contributions to the conductance from processes in which an electron tunnels into an empty or singly occupied level $\lambda$, respectively. Unlike the case without the Zeeman term, explicit summations over the magnetic quantum numbers remain, and we find
\begin{widetext}
\begin{eqnarray}
w_{\lambda }^{(0)} & = & \sum_{SM}b_{\lambda ,N,S}P_{N,S,M} \sum_{\sigma=\pm 1 }\left[ \frac{S + M \sigma +1}{2S+1 }f(\xi _{S+1/2,S,\sigma
}^{\lambda })+\frac{S-M \sigma}{ 2S+1}f(\xi
_{S-1/2,S,\sigma }^{\lambda })\right]\;, \label{wo} \\
w_{\lambda }^{(1)}& = &\sum_{S^{\prime }M^{\prime }}c_{\lambda ,N+1,S^{\prime
}}P_{N+1,S^{\prime },M^{\prime }} \sum_{\sigma=\pm 1}\left[ 
\frac{S^{\prime}+ M^\prime \sigma}{2S^{\prime}+ 1}f(-\xi _{S^{\prime },S^{\prime }-1/2,\sigma}^{\lambda })+\frac{S^{\prime }-M^{\prime }\sigma +1}{
2S^{\prime}+1}f(-\xi _{S^{\prime },S^{\prime }+1/2,\sigma}^{\lambda })\right] \;.\label{w1}
\end{eqnarray}
\end{widetext}
 Here $P_{N,S,M}= e^{-\beta [F_{N,S} + U_{N,S,M}]}/Z$ is the
probability of finding the dot with $N$ electrons, spin $S$ and spin 
projection $M$, with $F_{N,S}$ being the free energy of $N$ non-interacting electrons
 with total spin $S$, and 
$\!U_{N,S,M} \!= \!e^2 N^2/2C - J_s S(S+1) + g \mu_B B_\parallel M - \tilde \epsilon_{\rm F} N$. The Fermi-Dirac distribution $f(\xi _{S, S^{\prime },\sigma }^{\lambda })$ is evaluated at $\xi _{S,S^{\prime },\sigma }^{\lambda} = \epsilon_\lambda -J_{S}\left( S^{\prime }(S^{\prime }+1)-S(S+1)\right)-\tilde{\epsilon}_{\rm F}+g\mu_B B_\parallel\sigma/2 $ and $\sigma/2 =M^{\prime }-M$ is the spin projection of the electron that tunnels into the dot.  $\tilde{\epsilon}_{\rm F}=\epsilon_{\rm F}+e C_{\rm g}V_{\rm g}/C$ is an effective Fermi energy where $V_{\rm g}$ is the gate voltage and $C_{\rm g}$ the dot-gate capacitance. The coefficients $b_{\lambda ,N,S}=\left\langle \frac{1}{2}(\hat{n}_{\lambda }-1)(\hat{n}
_{\lambda }-2)\right\rangle _{N,S}$ and $c_{\lambda ,N+1,S^{\prime
}}=\left\langle \frac{1}{2}\hat{n}_{\lambda }(\hat{n}_{\lambda
}-1)\right\rangle _{N+1,S^{\prime }}$ are spin- and particle-projected quantities whose explicit expressions are given in Ref.~\onlinecite{RefThomas}.

\begin{figure}[h!]
\begin{center}
\epsfig{file=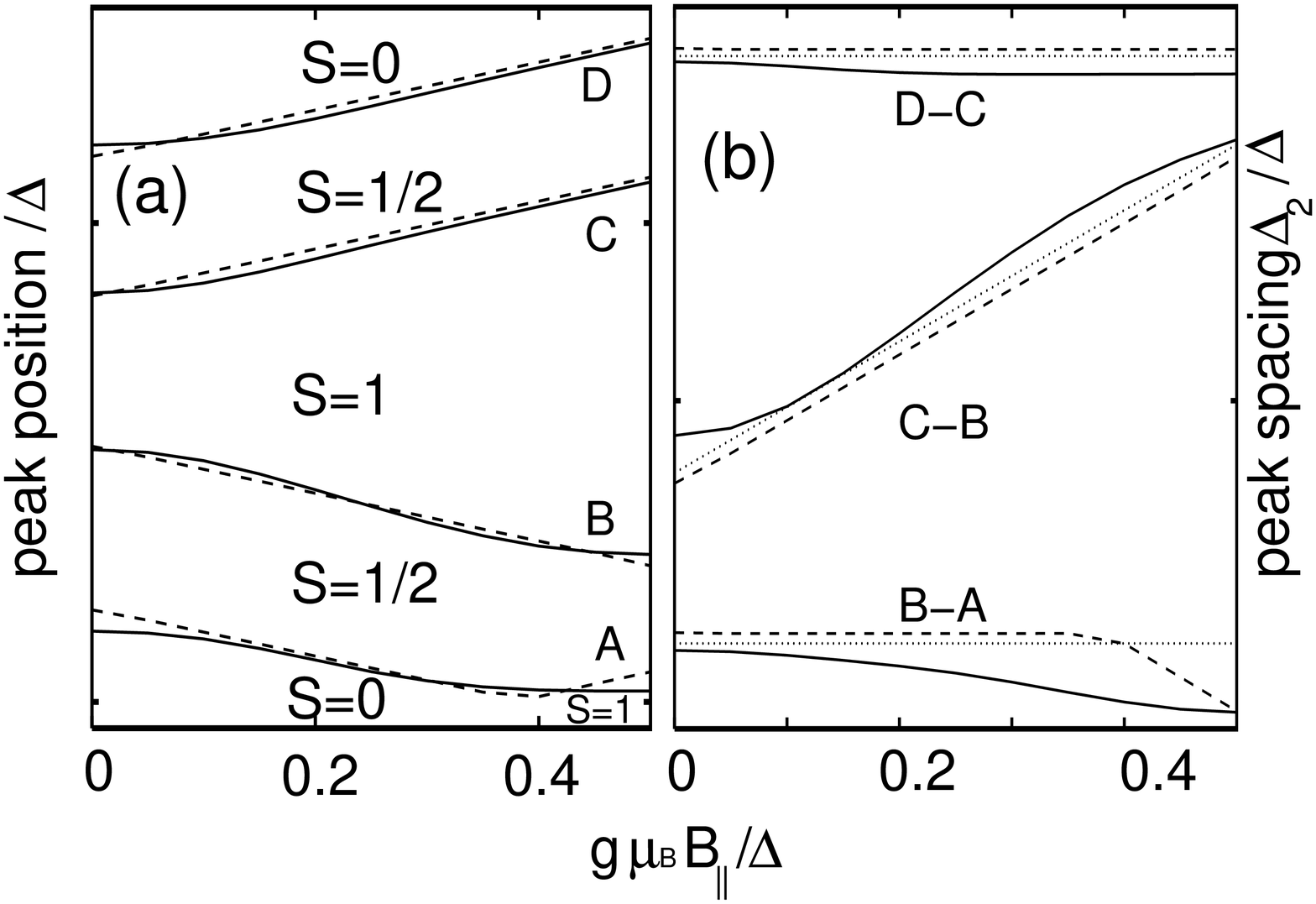,width=8.0 cm, clip=}
\end{center}
\caption{(a) CB peak position versus $g\protect\mu_B B_\parallel/\Delta$ for $
J_{S}=0.3\,\Delta$ and four consecutive peaks at $kT=0.01\,\Delta$ (dashed lines), and  $kT=0.1\,\Delta$ (solid
lines). (b) CB peak spacing versus $g\protect\mu_B B_\parallel/\Delta$ for the peaks shown in (a). The dotted lines are the
expected slopes of $\pm g\mu_B$ and $0$.}
\label{fig1}
\end{figure}

Using a realization of the single-particle Hamiltonian, we calculate the conductance from Eqs.~(\ref{cond}), (\ref{wo}) and (\ref{w1}) as a function of $\tilde\epsilon_{\rm F}$, and find its maximum to determine the peak position.  In Fig.~\ref{fig1}(a) we show calculated peak position versus $g \mu_B B_\parallel/\Delta$ for four consecutive peaks (A, B, C, D) and  for a realistic exchange constant of $J_{S}=0.3\,\Delta$ \cite{RefThomas}.  At $k T=0.01\,\Delta $ (dashed lines) the expected slopes of $\pm g\mu_B/2$ are observed.  Assuming a starting value of $S=0$ below peak A and using the observed peak position slopes, we assign GS spin values of $S = 0,
1/2, 1, 1/2, 0$ as electrons are added to the dot. At the higher temperature of $kT =0.1 \,\Delta$ (solid lines) the peak position curves become flat near $B_\parallel=0$ (i.e., for $g\mu_B B_\parallel \alt kT$) but the low-temperature slopes can still be identified. 

   Fig. \ref{fig1}(b) shows the corresponding peak spacings versus $g \mu_B B_\parallel/\Delta$.  Peak positions with alternating slopes lead to  peak spacings with  slopes of $\pm g \mu_B$. However,  two parallel consecutive peak positions result in a zero-slope peak spacing (e.g., $B-A$) and indicate that the GS spin increases (or decreases) twice in a row,  e.g., $0 \to 1/2 \to 1$. We see that the peak-spacing slopes do not change much at the higher temperature of $kT =0.1 \,\Delta$ and thus GS spins can be correctly inferred at this temperature.

\begin{figure}[t]
\begin{center}
\epsfig{file=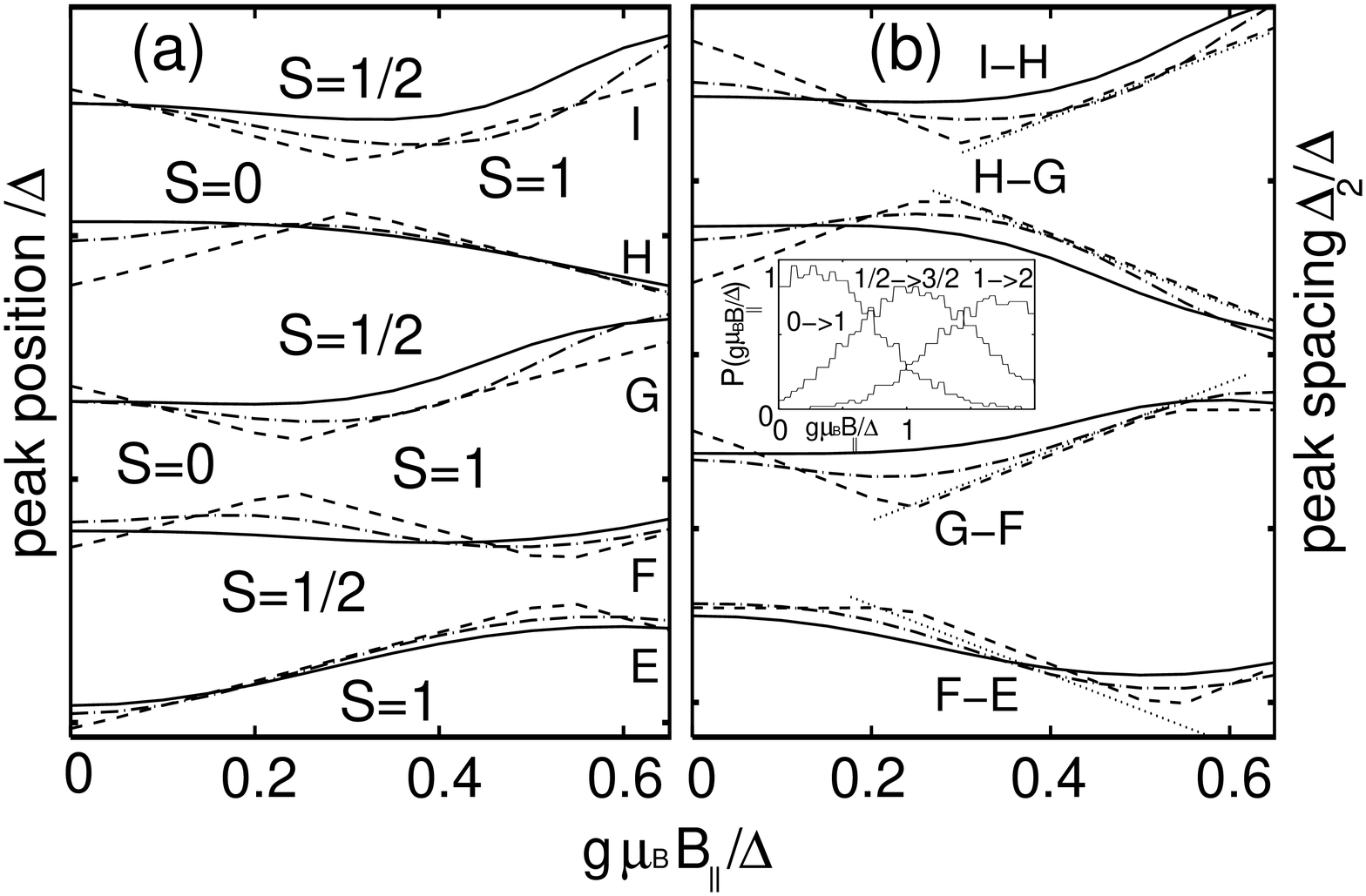,width=8.2cm, clip=}
\end{center}
\caption{(a) CB peak position versus $g\protect\mu_B B_\parallel/\Delta$ for 
$J_{S}=0.3 \,\Delta$ and for 5 consecutive peaks at  temperatures of $kT=0.01\,\Delta$
(dashed lines), $kT=0.1\,\Delta$ (dash-dotted lines), and $kT=0.15\,\Delta$
(solid lines). (b) CB peak spacing versus $g\protect\mu_B B_\parallel/\Delta$ for the peaks shown in (a). The dotted lines describe slopes of $\pm g\mu_B$. Inset:  the probability
distributions for $0 \rightarrow 1$, $1/2 \rightarrow 3/2$ and $1 \rightarrow 2$ crossings to occur at parallel field $B_\parallel$ and for $J_{S}=0.3\,\Delta$.}
\label{fig2}
\end{figure}

Fig.~\ref{fig2}(a) shows peak position versus $g \mu_B B_\parallel/\Delta$ for a
different set of five consecutive peaks (E, F, G, H, I) for $J_{S}=0.3\,\Delta $
and temperatures of $kT=0.01\,\Delta $ (dashed lines), $kT=0.1\,\Delta $ (dash-dotted lines)
and $kT=0.15\,\Delta $ (solid lines). At $kT =0.01\,\Delta$, level crossings are clearly observed in the form of kinks in the peak position versus $B_\parallel$.  For example, the dot between peaks H and I has $S=0$ in its zero-field GS, but at $g\mu_B B_\parallel/\Delta \sim 0.3$, an $S=1$ excited state crosses the $S=0$ state and becomes the new GS. The slope of the peak position versus $B_\parallel$ is still given by $g\mu_B \Delta S$, but now $\Delta S$ describes the change of the spin involving the new GS of the dot in which the level crossing occurs.  Thus the $g\mu_B/2$ slope of peak position H ($1/2\to 0$ transition) changes into a $-g\mu_B/2$ slope ($1/2\to 1$ transition) following the level crossing at $g\mu_B B_\parallel/\Delta \sim 0.3$.  Signatures of level crossings are also seen in the peak spacings of Fig.~\ref{fig2}(b) at $kT=0.01\,\Delta$.

 However, at the higher (but still low) temperatures of
$kT=0.1 \,\Delta$ and $kT=0.15 \,\Delta$, signatures of level crossings have almost completely disappeared.  We observe that when a level crossing occurs at a low value of $B_\parallel$, the peak position and peak spacing at 
$kT=0.15\,\Delta$ are essentially flat up to the crossing point. Flat slopes were observed at low fields for weakly coupled dots  in  Refs.~\cite{RefCMMarcus} and \cite{RefQD} (see Fig.~9 and Fig.~5, respectively) and are likely the result of temperature effects and/or level crossings.
 If the peak position and peak spacing are flat up to the crossing point, it is not possible to identify the slopes that are necessary for the correct GS spin identification. At higher values of the field, the peak spacing acquires a slope but this slope reflects the GS spin of the dot after the crossing. If such slopes are used to determine the zero-field GS spin, the spin value can be misassigned. For example, the dot between peaks H and I can be assigned a spin of $S=1$ instead of its true GS spin of $S=0$. 

 Spin misassignment is likely when a level crossing occurs at relatively low values of the field, reflecting a zero-field low-lying excited state  with spin $S^* > S$. The probability distributions for various level crossings to occur at a field $B_\parallel$ are shown in the inset of Fig.~\ref{fig2}(b) for $J_s=0.3 \,\Delta$.  In particular, the probability of a $0\to 1$ spin crossing at low $B_\parallel$ is quite substantial ($\sim 0.3$ for $g \mu_B B_\parallel/\Delta \alt 0.3$).  Thus the occurrences of $S=1$ ground states are likely to be overestimated. 

To avoid spin misassignment, we propose an improved method to determine the 
GS spin of the dot from parallel field measurements at temperatures $k T \alt 0.15\,\Delta$. The method is based on the observation that at these temperatures the conductance is dominated by the contribution from two groups of transitions between the $N$-- and $(N+1)$--electron dots, obtained by considering one level with spin $S$ in one of the dots and two levels with spins $S_0=S-1/2$ and $S_1=S+1/2$ in the other dot (e.g., $1/2 \to 0$ and $1/2\to 1$ transitions). In this case, the rate equations can be solved explicitly for any residual interactions \cite{RefTGY}. Assuming the relevant transitions from the $N$-- to the $(N+1)$--electron dots are $S \to S_0$ and $S \to S_1$ with reduced tunneling widths $\tilde\Gamma_0$ and $\tilde\Gamma_1$, respectively,  we find 
\begin{widetext}
\begin{equation} \label{2TM}
G=\frac{e^{2}}{\hbar k T}\sum_{M}P_{SM}^{(N)}\sum_{\sigma=\pm 1 }\left[ \frac{S-M\sigma}{\left( 2S+1\right)\left( 2S_{0}+1\right)}f\left(\xi _{0 \sigma}\right) \frac{\tilde\Gamma _{0}^{\rm l}\tilde\Gamma _{0}^{\rm r}}{\tilde\Gamma _{0}^{\rm l}+\tilde\Gamma_{0}^{\rm r}}+\frac{S + 1 +M\sigma}{\left( 2S+1\right)\left( 2S_{1}+1\right) }f\left(\xi_{1 \sigma}\right)\frac{\tilde\Gamma _{1}^{\rm l}\tilde\Gamma _{1}^{\rm r}}{\tilde\Gamma _{1}^{\rm l}+\tilde\Gamma_{1}^{\rm r}}\right]\;.
\end{equation}
\end{widetext}
Here $\xi_{i\sigma}= \varepsilon^{(N+1)}_{S_i} - \varepsilon^{(N)}_{S} -\tilde\epsilon_{\rm F}+ g\mu_B B_\parallel \sigma/2$ ($i=0,1$), where $\varepsilon^{(N)}_S$ is the many-particle energy of the level with spin $S$ in the $N$--electron dot, and $P^{(N)}_{S M}$ is the probability to find the dot in its $N$--electron state with spin S and spin projection $M$.  A similar expression can be derived when the relevant $N \to N+1$ transitions are $S_0 \to S$ and $S_1 \to S$.  This two-transition model takes into account the possibility of level crossing (versus $B_\parallel$) in the $N$-- or $N+1$--electron dot. The conductance formula (\ref{2TM}) is valid for arbitrary residual interactions, but is a low-temperature approximation $kT \alt 0.15 \,\Delta$. On the other hand, the conductance given by Eqs.~(\ref{cond}), (\ref{wo}) and (\ref{w1}) is only valid for the universal Hamiltonian but is exact for arbitrarily large temperatures.

\begin{figure}[h]
\begin{center}
\epsfig{file=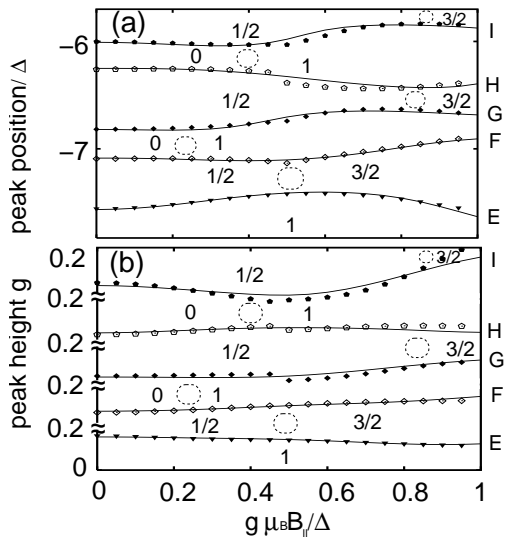,width=6.6 cm, clip=} 
\end{center}
\caption{Results of the two-transition model Eq.~(\ref{2TM}) (symbols) are fitted to the full calculation data of Fig. \ref{fig2} (solid lines) for both peak positions (panel (a)) and peak heights (panel (b)). 
Simultaneous fitting of both peak positions and
peak heights is needed to properly extract the values of the total spin from the data. Circles indicate crossing positions. }
\label{fig3}
\end{figure}

To determine unambiguously the GS spin, we fit simultaneously the measured peak position and peak height data versus parallel field using the above two-transition model. In a given range of $B_\parallel$, we assume a certain spin $S$ (for either the 
$N$-- or the $(N+1)$--electron dot) and fit four parameters: the two conductances $G_i\propto \tilde \Gamma^{l}_i \tilde \Gamma^{r}_i/(\tilde \Gamma^{l}_i + \tilde \Gamma^{r}_i)$ and the two many-body energy differences (e.g., $\varepsilon^{(N+1)}_{S_i} - \varepsilon^{(N)}_{S}$ if $S$ is the spin of the $N$--electron dot).  

 We validate our method using as data sets the $kT=0.15\,\Delta$ peak positions of Fig. \ref{fig2}(a) and their corresponding peak heights,  shown by the solid lines in Fig.~\ref{fig3}(a) and \ref{fig3}(b), respectively. To emulate the experimental situation, we treat the spins values and the crossing positions as unknown.  The fit of these data sets to the two-transition model (symbols in Fig.~\ref{fig3}) is found to be good. In particular, the corresponding spin values and crossing positions (dashed circles in Fig.~\ref{fig3}(a) and (b)) can be extracted systematically at a temperature of $k T=0.15\,\Delta $. 
We have also verified that the parameters we extract from the fit agree with their input values in the universal Hamiltonian. While it is possible to fit separately the peak position and peak height data, the best-fit parameters are often found to be unphysical. 

 Our method can also be used to extract the lowest energy level (at zero field) for a given spin value that is higher than the GS spin, as long as this level becomes the GS of the dot at some finite parallel field. 
 
In conclusion, we have derived a closed expression for the conductance in a parallel magnetic field of a CB quantum dot that is described by the universal Hamiltonian. For typical low temperatures of $k T\gtrsim
0.1\,\Delta$ used in the GaAs quantum dots, signatures of level crossings in a parallel field can be almost
completely washed out. When such a crossing occurs at low values of the field, the GS spin of the dot, as determined from the peak position motion in a parallel field, can be misidentified.
 We have described an improved  method in which the measured peak position and peak height data are fitted to a finite-temperature two-transition model that takes into account level crossings.   The
model is valid for any interaction in the dot, and thus could be generally used as to extract the GS spin in experimental situations where the lowest attainable temperature is $\sim 0.1 \,\Delta$. 

This work was supported in part by the U.S. DOE grant DE-FG-0291-ER-40608. We thank
B.L. Altshuler, Ya.M. Blanter, P.W. Brouwer, M. Eto, C.M. Marcus, P.L. McEuen, and Yu.V. Nazarov for useful discussions. 
D. H-H is grateful to the Theoretical Physics Group of the Kavli Institute of Nanoscience Delft for its hospitality.

\end{document}